# Enhanced Planar Antenna Efficiency Through Magnetic Thin-Films

Zhi Yao, *Member, IEEE*, Sidhant Tiwari, *Member, IEEE*, Joseph Schneider, *Member, IEEE*, Robert N. Candler, *Senior Member, IEEE*, Gregory P. Carman, and Yuanxun Ethan Wang, *Fellow, IEEE*

*Abstract*— **This work proposes to use magnetic material as the substrate of planar antennas to overcome the platform effect caused by the conducting ground plane. The upper bound of the radiation efficiency of an electric-current-driven low-profile antenna is theoretically derived, which is inversely proportional to the Gilbert damping factor of the magnetic material. Meanwhile, the improvement of radiation due to the use of magnetic material is demonstrated by a three-dimensional (3D) multiphysics and multiscale time-domain model. The simulation results match the theoretical derivation, showing 25% radiation efficiency from a planar antenna backed by a FeGaB thin film with 2.56 μm thickness. Furthermore, for conductive ferromagnetic materials, it is shown that the eddy current loss can be well suppressed by laminating the thin film into multiple layers. The radiation efficiency of the modeled antenna with a conductive ferromagnetic substrate is improved from 2.2% to 11.8% by dividing the substrate into 10 layers, with a ferromagnetic material fill factor of 93%.**

## I. INTRODUCTION

Scaling down of circuitry has been a growing trend in modern electronics, enabling miniaturized and interconnected systems. Specifically, conformal devices with very small thicknesses are popular in applications such as wearable devices for law enforcement, military, and civilian emergency services [1], [2]. However, a major challenge that stands in the way of realizing these new technologies is scaling down the antenna. Planar dimensions of traditional antennas must be on par with the electromagnetic (EM) wavelength to transmit efficiently [3]. Moreover, space-saving low-profile antennas require a ground plane for operation, causing poor radiation due to the platform effect. The platform effect is the major issue that conventional current-based antennas suffer when placed at a short distance above a conducting plane. The radiation becomes inefficient because the image current flows in the opposite direction and cancels the original current source. Therefore, antenna scaling is prevented by the excessive reactive energy stored between the radiating element and the conducting plane, thus raising the radiation quality factor ($Q$ factor) and making the antenna difficult to match. In order to alleviate the platform effect and to allow for miniaturization, new materials and technologies need to be implemented.

One of the effective ways to increase efficiency is to replace the regular dielectric substrates with magnetodielectric materials, which provide high values of both relative permittivity $\epsilon_r$ and relative permeability $\mu_r$. In patch antennas with such substrates, the effective EM wavelength is reduced by approximately $\sqrt{\epsilon_r \mu_r}$ times, leading to a reduction of antenna characteristic length by approximately the same scale factor [4]. Efforts have been made on implementing both natural ferrites [5]–[7] and artificial magnetic materials, such as metamaterials [8]. When the material loss is considered, modeled with a the complex permeability $\mu_r = \mu_r' - j\mu_r''$, the common understanding in these works is to avoid large values of $\mu_r''$, which in turn limits the value of $\mu_r'$ to be below a hundred and the operation frequency of the natural magnetic materials to be below hundreds of megahertz [9], i.e. below domain wall resonance frequencies. However, as the frequency increases to the gigahertz range, magnetic materials now exhibit ferromagnetic resonance (FMR), manifesting itself as dramatically high values of $\mu_r''$ resulting in large values for the magnitude of $\mu_r$. Moreover, there is a causality relation between $\mu_r'$ and $\mu_r''$, meaning these two values cannot be tuned independently of each other. The ferrites/ferromagnets perform as imperfect magnetic conductors, converting the electric current image [10], [11] into one that is parallel to the original source current, enhancing the radiation from the source rather than canceling it. The strong magnetic flux existing in the antenna structure enables the antenna to switch from being electric field dominated to magnetic field dominated. Following this strategy, new types of radiating mechanisms targeting antenna miniaturization have been proposed. These

Manuscript received **, 2021. The work was supported by NSF Nanosystems Engineering Research Center for Translational Applications of Nanoscale Multiferroic Systems (TANMS) Cooperative Agreement Award (No. EEC-1160504), and the Defense Advanced Research Projects Agency (DARPA) Magnetic Miniaturized and Monolithically Integrated Components (M3IC) Program under award W911NF-17-1-0100.

Zhi Yao is with Computational Research Division, Lawrence Berkeley National Laboratory, Berkeley, CA 94720, USA (email: jackie_zhiyao@lbl.gov). Sidhant Tiwari is with Sandia National Laboratory, Albuquerque, NM 87123, USA. Joseph Schneider is with Lawrence Livermore National Laboratory, Livermore, CA 94550, USA. Robert N. Candler and Yuanxun Ethan Wang are with the Department of Electrical and Computer Engineering, University of California, Los Angeles, CA 90095, USA. Robert N. Candler is jointly with the California NanoSystems Institute (CNSI), Los Angeles, CA 90095, USA. Gregory P. Carman is with the Mechanical and Aerospace Engineering Department, University of California, Los Angeles, CA 90095, USA.

include a strain-mediated antenna composed of composite multiferroic materials [12], [13] and mechanical antennas based on physically oscillating magnets [14], [15]. By utilizing the time-varying magnetic flux as the radiating source, these new types of antennas could potentially be immune to the conductive loss and the platform effect. The recent work on electrically small loop antenna with a ferrite substrate has experimentally demonstrated the enhancement of the antenna radiation performance by FMR of the ferrite [16]. It is worth pointing out that the multiferroic antenna and mechanical antenna both rely on high permeability magnetic material to lower the radiation quality factor, which is consistent with the approaches attempted by merely using magnetodielectric substrates [5]–[8].

The magnetic material used in magnetic antennas should possess an FMR frequency in the gigahertz range, large value of permeability, as well as low eddy current loss, i.e. low electric conductive loss. To achieve gigahertz FMR frequency and large permeability at the same time, in-plane biased magnetic thin films should be used. The reason is that, compared to bulk materials, the FMR frequency of an in-plane biased thin film is increased by the factor of $\sqrt{M_S/H_0}$, where $M_S$ is the saturation magnetization and $H_0$ is the magnetic DC bias. As the FMR frequency is increased in the in-plane biased case, lower bias magnetic fields can be used to achieve the same FMR frequencies, easing the requirements on the strength of the electromagnets used to provide the bias magnetic fields. Moreover, rather than ferrites, ferromagnetic materials are preferred. The reason is that in saturated ferromagnetic materials, all the spins are aligned to the bias direction, leading to a large spontaneous magnetization and a higher permeability near FMR. This is in contrast to the saturated ferrites, where adjacent spins are opposite to each other and different in magnitude, leading to a low saturation magnetization. Such spin orientation results in a smaller net magnetization thus lowering the magnitude of the permeability near the FMR frequency. However, even though ferromagnetic materials provide large permeability, they are typically conductive, leading to severe eddy current loss. Therefore, one also needs to resolve the dilemma between having the large magnetic permeability from ferromagnetic materials and requiring low conductive losses. It is proposed in this work to suppress the eddy current loss by segmenting the ferromagnetic film into thin layers, so that the giant eddy current loops are broken, and the conductive loss is reduced. In summary, this work proposes that by inserting in-plane biased, multi-layered, ferromagnetic thin films between the radiating source and the conducting ground plane, one can drastically improve the radiation performance, such as radiation efficiency.

In this manuscript, the elimination of the platform effect is demonstrated by studying a planar antenna. The antenna is composed of a planar electric current backed by a ferromagnetic thin film. At the bottom of the thin film, a perfect electric conducting (PEC) ground plane is assigned. The radiation efficiency of this idealized radiator is derived theoretically, based on the plane wave assumption. Furthermore, a three-dimensional (3D) finite difference time domain (FDTD) algorithm is developed [17], [18], to demonstrate the theory. The algorithm is based on the alternating directional implicit (ADI) method to achieve unconditional stability. In the model, an electric current source on top of an FeGaB substrate is used. The modeling results show that even if the substrate is only several-micrometer-thick, it can boost up the radiation power by six orders of magnitude. Additionally, the simulation results demonstrate the suppression of eddy current loss by laminating the continuous thin film into multiple layers. With laminated ferromagnetic substrate, the radiation efficiency of the current source with the ferromagnetic substrate can be improved from 2.2% to 12% by dividing the substrate into 10 layers.

## II. THEORY

Consider an infinite, uniform current sheet $i_S$ that radiates EM waves into free space, as shown in Fig. 1(a). The current source is placed over an infinite PEC ground plane, with a ferromagnet substrate inserted between them. The PEC-backed current source is the model of the antenna. The thickness of the substrate is electrically small such that $kh \ll 1$, where $k$ is the wave number in the substrate. The ferromagnetic substrate is biased to saturation by an in-plane magnetic DC bias $H_0$ that is parallel to the current source $i_S$. EM waves are directly radiated into the free space and reflected by the PEC plane, resulting in destructive interference. Therefore, according to the classical EM theory, the amplitude of the time-varying EM waves in the different regions are:

Stored field: 
$$\begin{cases} E_y = E_0 \sin(kz) \\ H_x = -E_0/j\eta \cos(kz) \end{cases},$$

Radiated field: 
$$\begin{cases} E_y = E_0 \sin(kh) \exp(-jk_0 z) \\ H_x = -E_0/\eta_0 \sin(kh) \exp(-jk_0 z) \end{cases}, \quad (1)$$

where $E_0$ is the aperture electric field amplitude at the interface between the free space and the substrate.

The radiated power into the free space is thus calculated as

$$P_{rad} = \frac{1}{2\eta_0} \iint_S |E|^2 \, ds \approx \frac{1}{2\eta_0} E_0^2 (kh)^2 S$$
$$= \frac{1}{2\eta_0} E_0^2 h^2 S \omega^2 |(\mu' - j\mu'')\epsilon|, \quad (2)$$

Note that in Equation (2), the off-diagonal permeability terms in the Polder tensor are ignored for mathematical simplicity. In Equation (2), the approximation of a linear distribution along the z-direction for the electric field is applied since the tangential electric field on the PEC is zero. Similarly, an approximation of a uniform magnetic field distribution along the z-direction leads to the stored magnetic energy and magnetic power loss as in Equations (3) and (4), respectively.

$$W_H = \frac{1}{2} \iiint_{z<h} \mu' |H|^2 \, dv \approx \frac{1}{2} \mu' |H|^2 hS = \frac{1}{2} \mu' \frac{E_0^2}{\left| \sqrt{\frac{(\mu' - j\mu'')}{\epsilon}} \right|^2} hS,$$
$$(3)$$

$$P_M = \frac{1}{2}\,\omega \iiint_{z<h} \mu'' |H|^2\, dv \approx \frac{1}{2}\,\omega \mu'' \frac{E_0^2}{\left|\sqrt{\frac{(\mu'-j\mu'')}{\epsilon}}\right|^2}\, h, \qquad (4)$$

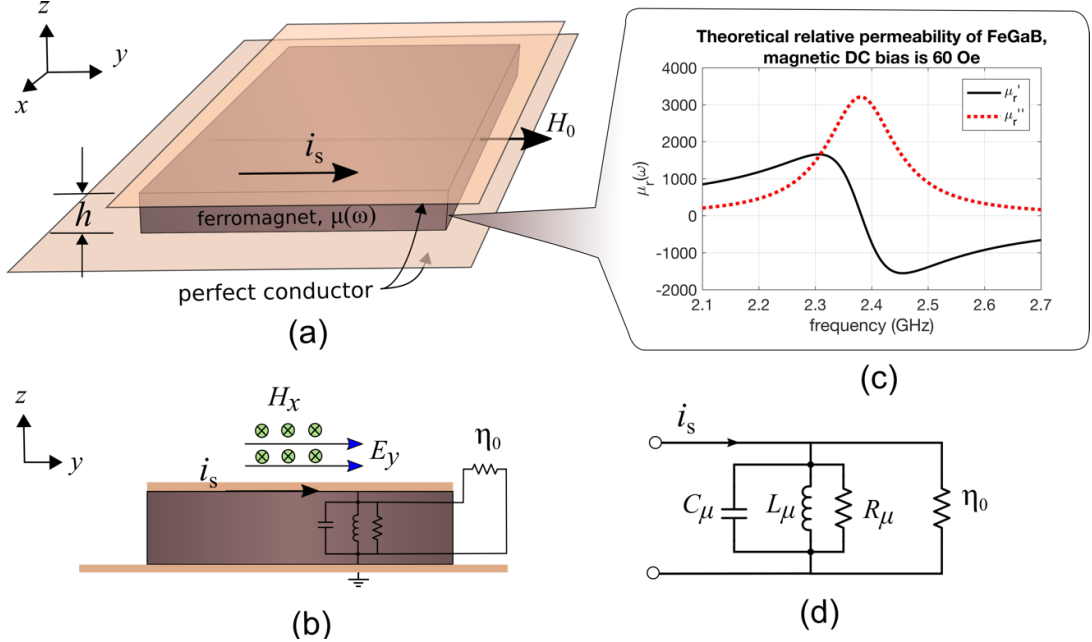

Fig. 1. Radiation from electric current source. The coordinate system is chosen such that the surface current is in the y direction. The thickness of the substrate satisfies the condition $kh \ll 1$. (a) An infinite, uniform current sheet $i_S$ that radiates EM waves into free space. The current source is grounded by an infinite perfect electrically conducting (PEC) plane, with a ferromagnet substrate inserted between them. The thickness of the substrate is electrically small such that $kh \ll 1$, where $k$ is the wave number in the substrate. The ferromagnetic substrate is biased to saturation by an in-plane magnetic DC field $H_0$ that is aligned with the current source $i_S$. Note that in the numerical model, the actual size of the structure is set to be 5×5×10 μm in the $x$, $y$, and $z$ directions, respectively. The infinite planar size is realized by periodic boundary conditions applied at the four side walls. (b) The $y$-$z$ cross section of (a). (c) Analytical permeability spectrum of a ferromagnetic material, FeGaB, under a DC magnetic bias of 60 Oe, with the saturation magnetization $4\pi M_S$ being $1.2 \times 10^4$ Gauss. (d) Circuit model of the structure in (a) and (b). The parallel RLC resonator represents the ferromagnetic resonance, and the shunt resistor $\eta_0$ represents the intrinsic resistance of the fee space.

In ferromagnetic materials, the electric energy stored in the structure is negligible compared to the magnetic stored energy, thus the total amount of stored energy is approximately equal to the magnetic stored energy, or in the mathematical form, $W_{\text{total}} \approx W_H$. Therefore, the total quality factor of the system is given by:

$$Q_{total} \approx \omega \frac{W_H}{P_{rad}+P_M} = \frac{1}{\frac{h\omega}{\eta_0}\frac{\mu'^2+\mu''^2}{\mu'}+\frac{\mu''}{\mu_r}} = \frac{\mu_r'}{hk_0\left(\mu_r'^2+\mu_r''^2\right)+\mu_r''}. \qquad (5)$$

Note that in Equation (5), the electric power loss is neglected by assuming that the conductivity of the material is zero. Similarly, the radiation quality factor is

$$Q_{rad} \approx \omega \frac{W_H}{P_{rad}} = \frac{1}{\frac{h\omega}{\eta_0}\frac{\mu_r'^2+\mu_r''^2}{\mu'}} = \frac{\mu_r'}{hk_0\left(\mu_r'^2+\mu_r''^2\right)}. \qquad (6)$$

Hence, for an antenna working around 2 GHz, $Q_{\text{rad}}$ is on the order of $10^4$ when the thickness of the substrate is 1.5 μm if the material is non-magnetic, or in mathematical form, $\mu_r' = 1$ and $\mu_r'' = 0$. Since traditional antennas are mostly made of conductors and rely on conductive current to radiate, the platform effect is an inevitable shortcoming of traditional low-profile antennas. The platform effect results in more energy being stored in the structure instead of being radiated away into free space, raising the antenna quality factor. On the other hand, a magnetic substrate with high relative permeability offers the capability of significantly lowering $Q_{\text{rad}}$ and improving the radiation performance of low-profile antennas.

According to the definition, the radiation efficiency can be calculated as

$$\xi_{rad} = \frac{Q_{total}}{Q_{rad}} = \frac{1}{1+\frac{\mu_r''}{\mu_r'^2+\mu_r''^2}\frac{1}{k_0h}}. \qquad (7)$$

In saturated ferromagnetic materials that is biased in-plane, the relative permeability is calculated by (8) [19].

$$\mu_r = \frac{\omega_m}{\omega_0}\frac{\omega_r^2}{\omega_r^2-\omega^2+j\alpha\omega(2\omega_0+\omega_m)} + 1. \qquad (8)$$

In (8), $\omega_0$ is the Larmor frequency, defined as $\omega_0 = \mu_0\gamma H_0$. Similarly, $\omega_m$ is defined as $\mu_0\gamma M_S$, where $M_S$ is the saturation magnetization of the ferromagnet. The term $\omega_r$ stands for ferromagnetic resonance frequency, which is calculated according to Kittel's equation as $\omega_r = \mu_0\gamma\sqrt{H_0(H_0+M_S)}$ for in-plane biased films. The term $\alpha$ is the Gilbert damping constant of the ferromagnetic material, which is related to the FMR linewidth by the formula $\Delta H = 2\alpha\omega/\mu_0\gamma$. Substituting (8) into (7) yields [16]

$$\xi_{rad} = \frac{1}{1+\alpha\omega\frac{\omega_0}{\omega_m}\frac{2\omega_0+\omega_m}{\omega_r^2}\frac{1}{k_0h}} \ , \qquad (7')$$

Note that in the derivation of Equation (9), the latter term of Equation (8) (i.e. the constant number 1) is neglected. This is a valid approximation as near FMR, the value of $\mu_r''^2$ is very large such that $\mu_r'^2 + \mu_r''^2 \approx \chi_r'^2 + \mu_r''^2$.

However, as mentioned previously, ferromagnetic materials are electrically conductive. Therefore, heat dissipation will be generated by the oscillating electric field in such materials. With a conductivity $\sigma$ in the ferromagnetic film, the electric power dissipation can be calculated as in

$$P_E = \frac{1}{2}\iiint_{z<h}\sigma|E|^2\,dv \approx \frac{1}{2}\iiint_{z<h}\sigma E_0^2|kz|^2\,dv$$

$$= \frac{1}{6}\sigma E_0^2\omega^2\left|\sqrt{(\mu'-j\mu'')\epsilon}\right|^2 h^3 S, \qquad (9)$$

Similarly,

$$\xi_{rad} = \frac{Q_{total}}{Q_{rad}} = \frac{P_{rad}}{P_{rad}+P_M+P_E} = \frac{1}{1+\frac{\mu_r''}{\mu_r'^2+\mu_r''^2}\frac{1}{2k_0h}+\frac{\sigma h\omega\mu_0}{3k_0}} \ . \quad (10)$$

Substituting Equation (8) into Equation (10) yields:

$$\xi_{rad} = \frac{1}{1+\alpha\omega\frac{\omega_0}{\omega_m}\frac{2\omega_0+\omega_m}{\omega_r^2}\frac{1}{k_0h}+\frac{\sigma h\omega\mu_0}{3k_0}} \ , \qquad (11)$$

Equation (11) again reveals the previously mentioned conclusion that the electric conductivity of the substrate decreases the radiation performance of the structure.

## III. Model

In order to accurately simulate the performance of the low-profile antenna considered in this work, an algorithm that describes both the EM wave propagation and the micromagnetic dynamics is applied [20]. Mathematically, this algorithm simultaneously solves the Maxwell's Equations (12) and the Landau-Lifshitz-Gilbert (LLG) Equation (13) using finite-difference time-domain (FDTD) method in a coupled fashion simultaneously.

$$\nabla \times \boldsymbol{H} = \epsilon\frac{\partial \boldsymbol{E}}{\partial t} + \boldsymbol{J} + \sigma\boldsymbol{E}, \nabla \times \boldsymbol{E} = -\frac{\partial \boldsymbol{B}}{\partial t} \ , \qquad (12)$$

$$\frac{\partial \boldsymbol{M}}{\partial t} = \mu_0\gamma\big(\boldsymbol{M}\times\boldsymbol{H}_{eff}\big) + \frac{\alpha}{|\boldsymbol{M}|}\boldsymbol{M}\times\frac{\partial \boldsymbol{M}}{\partial t}. \qquad (13)$$

In Equation (12), $\boldsymbol{E}$ represents the electric field, $\boldsymbol{J}$ represents the electric current volume density, $\boldsymbol{H}$ represents the magnetic field intensity, and $\boldsymbol{B}$ represents the magnetic flux density. The material properties are also involved in the Maxwell's equations, where $\sigma$ is the electric conductivity and $\epsilon$ is the relative permittivity. In Equation (13), $\mu_0$ represents the vacuum permeability. $\gamma$ is the gyromagnetic ratio, possessing a value of $-1.759\times10^{11}C/kg$. The term $\alpha$ is the Gilbert magnetic damping constant defined as $\alpha = \mu_0\gamma\Delta H/4\pi f_t$, with $\Delta H$ being the FMR linewidth and $f_t$ being the frequency at which the linewidth is measured. As the governing law of micromagnetics, the LLG Equation (13) describes the evolution of magnetization $\boldsymbol{M}$, with $\boldsymbol{H}_{eff}$ being the total effective magnetic field that drives the magnetic spins [21]. The low-profile antennas considered in this work consist of structures with characteristic dimensions much smaller than the EM wavelength. Conventional FDTD operates under the limit of Courant–Friedrichs–Lewy (CFL) stability condition, which incurs a tremendous amount of calculation, especially with such a small-scale structure. To overcome the stability constraint and reduce the time consumption of the simulation, alternating direction implicit (ADI) methods are used to obtain unconditional stability. The size of the entire simulation space shown in Fig. 1(a) is 5×5×10 μm in the $x$, $y$ and $z$ directions, respectively. The lower boundary, as mentioned, is set to be PEC. The ferromagnetic film is placed on the PEC, with the thickness being $h = 2.56$ μm, and the planar dimensions 5×5 μm. Periodic boundary conditions are applied at the four side walls to realize the infinite dimensions of the magnetic film in the planar directions. The upper surface of the space is terminated with absorbing boundary. A uniform electric current excitation is applied on the top surface of the ferromagnetic thin film. The field components are defined such that all the electric field components are along the edges of the spatial cell, and all the magnetic field components are face-centered on the cell surfaces. The spatial resolution is $\Delta x$=1 μm, $\Delta y$=1 μm, $\Delta z$=0.01 μm, and the time step is set as $\Delta t$=2.31×10$^{-13}$ s, which is $10^4$ times of the CFL limit. The surface current excitation center frequency is 2.4 GHz, and it is in the form of a modified Gaussian pulse, with a bandwidth of ±500 MHz. In this work, FeGaB is used due to its attractive ferromagnetic properties, e.g., low electric conductivity and high saturation magnetization. In the model, the saturation magnetization of FeGaB is $4\pi M_S = 12000$ Gauss, the FMR linewidth is $\Delta H = 30$ Oersted , and the electric conductivity is $\sigma = 5\times10^5$ Siemens/meter [22]. The magnetic DC bias applied in-plane is 60 Oersted so that the FMR frequency and the input signal frequency overlap.

Fig. 2 shows the simulated results with the geometry and material properties specified in the last paragraph. To explore the effect of conductive dissipation on the radiation performance, two additional cases have been simulated:

1. Artificial nonconductive ferromagnetic substrate: The setup of this control case is identical to the one previously introduced, except that the conductivity of FeGaB is artificially set to be zero. Therefore, this case is the optimal circumstance with a nonconductive ferromagnetic material.
2. No substrate under the current. In this case, the space between the current source and the PEC ground is filled with air, which is the original platformed antenna with the current source close to the PEC.

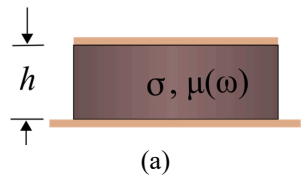

(a)

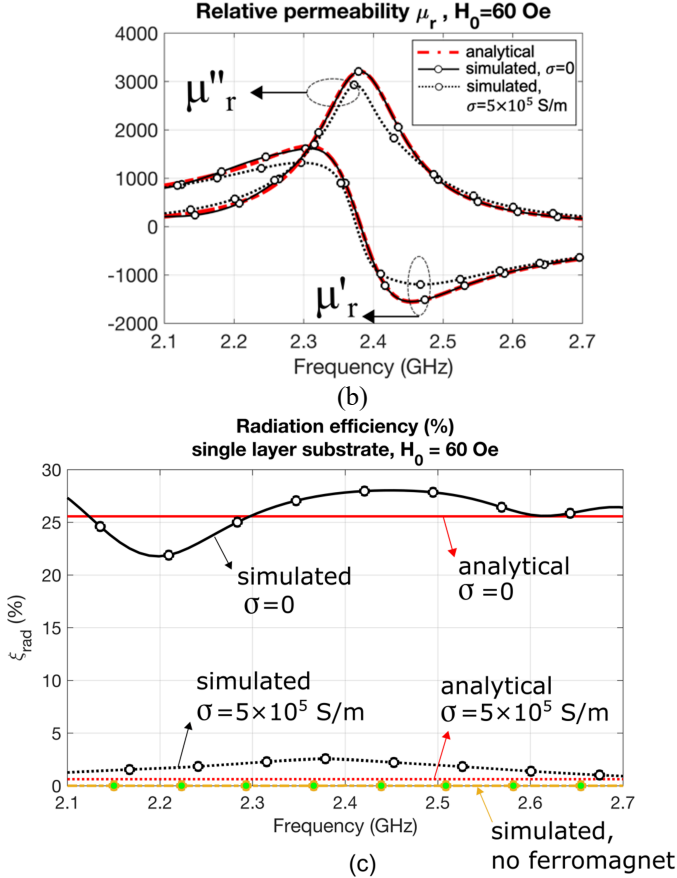

Fig. 2. Simulated radiation efficiency from the electric current source. (a) Antenna structure. The current is placed on the top surface of the ferromagnetic substrate. The size of the entire simulation space is 5×5×10 μm in the x, y and z directions, respectively. The spatial resolution is Δx=1 μm, Δy=1 μm, Δz=0.01 μm. (b) Simulated permeability of the FeGaB thin film, with conductivity of 5 × $10^5$ Siemens/meter , and an artificial conductivity of zero. (c) Simulated radiation efficiency with conductive and nonconductive materials, as well as without substrate.

As can be seen in Fig. 2(b), the simulated permeability is almost independent from the electric conductivity, and it matches the analytical results. At FMR, the imaginary permeability $\mu''$ is as large as 3000, and close to FMR the real permeability $\mu'$ is as large as 1500. Fig. 2(c) shows the comparison between the radiation efficiency of different substrate materials. Without the ferromagnetic substrate, the simulated radiation efficiency is on the order of $10^{-7}$ (as shown by the yellow curve with triangle marks), indicating that almost the entire radiated field is cancelled out by the PEC platform. Compared to the air-filled antenna, the artificial nonconductive ferromagnetic substrate improves the radiation efficiency by $10^6$ times, leading to a radiation efficiency of 25%, shown by the black solid curve with circle marks. The red solid line represents the analytical radiation efficiency calculated by Equation (7'), showing a good match to the simulation. However, this is the ideal case with no eddy current loss. Practically, the conductive FeGaB results in the radiation efficiency being only 2.2%, as shown by the black dashed curve with circle marks in Fig. 2(c). The analytical radiation efficiency corresponding to the FeGaB material is calculated with Equation (11), and plotted as the red dashed curve in Fig. 2(c). Therefore, in order to achieve the full advantage of using a ferromagnetic substrate to improve the radiation performance, modified geometries need to be explored to suppress the eddy current loss.

## IV. EDDY CURRENT SUPPRESSION

It is evident that magnetic materials with high relative permeability help overcome the platform effect. However, unfortunately, most of the materials that have such high permeability are ferromagnetic materials, which are highly conductive and suffer significant eddy current loss. By briefly analyzing Faraday's law in the integral form $\oint_C \boldsymbol{E} \cdot d\boldsymbol{l} = -(\partial/\partial t) \iint_S \boldsymbol{B} \cdot d\boldsymbol{A}$, one can quickly conclude that the eddy current loss can be well suppressed by reducing the magnetic flux by laminating the thin film into multiple layers, as shown in the inset of Fig. 3. Note that the thickness of the laminates should be at least comparable to the skin depth, so that the eddy current loop could be broken into smaller loops and the conductive loss will be reduced.

Fig. 3 shows the radiation efficiency of antennas with laminated substrates of various numbers of layers and different layer thicknesses. The gap between the PEC ground and the current source is approximately 2.56 μm for each lamination geometry. Since the planar dimensions of the antenna structure are constant, the thickness ratio represents the volume ratio of the material. The skin depth of FeGaB close to FMR is approximately 0.3 μm. Therefore, laminates with thicknesses smaller than 0.3 μm are effective for the eddy current suppression, such as the 8-layer, 10-layer and 12-layer structures, leading to the peak radiation efficiency of 9.27%, 11.8% and 10.31%, respectively. It is noticed that a dispersive radiation efficiency spectrum is formed, in contrast to the single-layer cases simulated in Fig. 2, where the radiation efficiency is constant over the frequency band under observation. Peaks in the dispersive spectrum $\xi_{\mathrm{rad,max}}(f)$ are formed around 2.4 GHz, due to the FMR effect influenced by the inductive-capacitive coupling between the laminates.

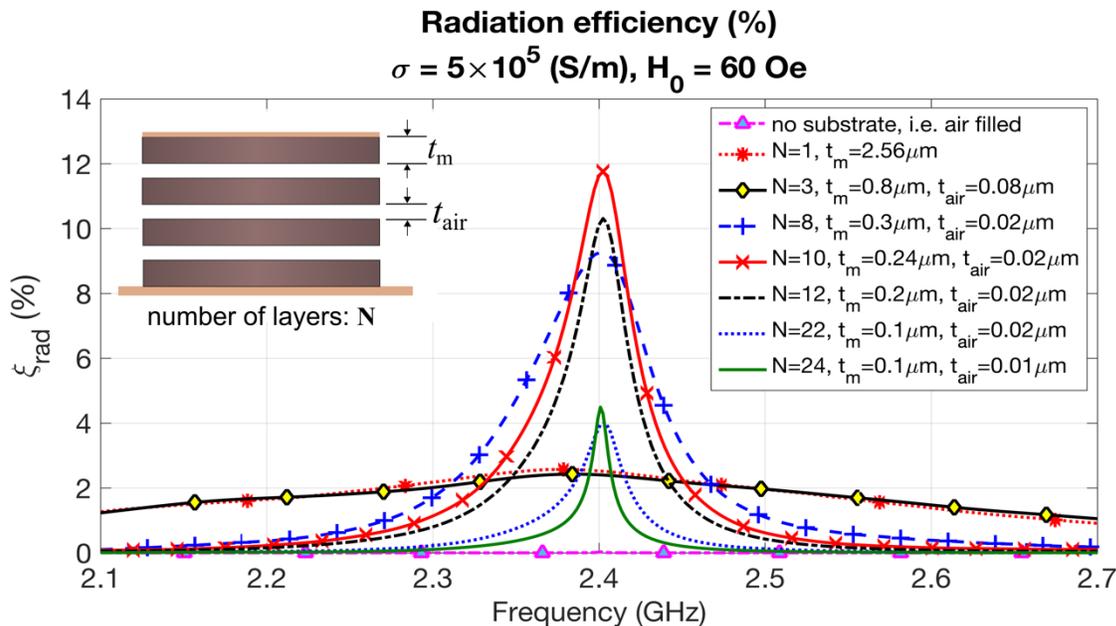

Fig. 3. Radiation efficiency of antennas with laminated substrate with various numbers of layers and different layer thicknesses. Inset: geometry of the laminated substrate. $t_m$: thickness of the ferromagnetic layers; $t_{air}$: gap between adjacent layers. N: number of ferromagnetic layers.

Intuitively, the more layers the ferromagnetic substrate is cut into, the more effectively the eddy current loops will be broken down, and the better suppression effect will be achieved. However, different from the intuition, it is observed from Fig. 3 that a maximum radiation efficiency is achieved with N=10 (N is the number of ferromagnetic layers), instead of the maximum number N=24. This optimum number of laminate layers is attributed to the balance between the interlayer coupling effects and the ferromagnetic material volume fraction. One type of possible interlayer coupling is capacitive coupling between adjacent ferromagnetic layers through the air gap, which is elaborated in Appendix A. To be brief, the electric fields couple to each other through the air gap, leading to an equivalent continuous, giant dielectric eddy current loop, which degrades the radiation efficiency. This effect grows stronger for thinner air gaps. Another type of interlayer coupling captured by the FDTD model is coupling via dipolar magnetic fields generated by the magnetization in each layer. Therefore, competing effects exist between the interlayer coupling, the ferromagnet/air volume ratio, and the change in the eddy current suppression effectiveness as the ferromagnetic layer thickness changes. This multi-factor scenario is captured by the numerical model proposed in this work, while is too complicated to be captured by theoretical analysis. In summary, the numerical model proposed in this work serves as a tool for optimizing the antenna structure for the purpose of suppressing the eddy current loss to the largest extent.

## V. CONCLUSION

In this work a layered ferromagnetic material is explored as a method to increase radiation efficiency in planar electrically small antennas. Air-filled planar electrically small antennas radiate almost no power into the space due to an image current that is anti-parallel to the source current. Furthermore, the radiation quality factor ($Q_{rad}$) is large due to the reactive energy associated with this antenna geometry. Inserting a ferromagnetic material between the source and ground plane enhances the radiation efficiency by reversing the direction of the image current in addition to reducing $Q_{rad}$ through the high relative permeability of the material. Radiation efficiency is further improved by dividing the ferromagnetic material into several layers reducing eddy current losses, thus, lowering the negative effects caused by the materials conductivity. An ADI-FDTD model is developed coupling the LLG Equation with Maxwell's Equations to study the complicated dynamics of a planar antenna with a laminated ferromagnetic substrate. Numerical results show that inserting FeGaB into the airgap increases efficiency from lower than 0.00001% to greater than 2%. Furthermore, dividing the FeGaB into 10 layers further improves the radiation efficiency to approximately 2% to 11.8%. These results show that the radiation efficiency of planar electrically small antennas can be dramatically increased by using a ferromagnetic material inserted between the source plane and the ground plane.

## ACKNOWLEDGMENT

The work was supported by NSF Nanosystems Engineering Research Center for Translational Applications of Nanoscale Multiferroic Systems (TANMS) Cooperative Agreement Award (No. EEC-1160504), and the Defense Advanced Research Projects Agency (DARPA) Magnetic Miniaturized and Monolithically Integrated Components (M3IC) Program under award W911NF-17-1-0100.

## APPENDIX A: CIRCUIT MODELLING OF DISPLACEMENT EDDY CURRENT FLOW

### A. Analysis Setup

As discussed in the main body of this work, a common method to reduce eddy currents in a ferromagnetic material is to laminate the material into several thin sheets separated by insulators. This has the effect of restricting the possible paths of the eddy currents, limiting the amount of current that flows and thus, limiting the power dissipated. However, as the layers get too thin, interlayer coupling can grow as the layer boundaries become closer together. One type of interlayer coupling is the capacitive coupling between the ferromagnetic laminations through the dielectric gaps.

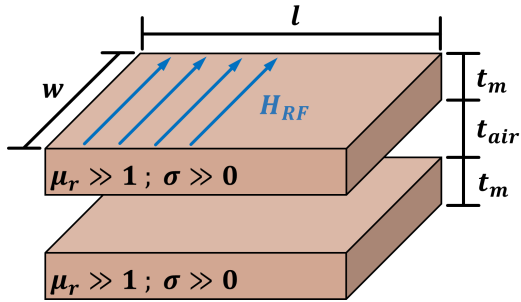

Fig. A1. Structure used to study the effect of displacement eddy currents on power dissipation: Two conductive ferromagnetic layers of thickness $t_m$ separated by an air gap of $t_{air}$. A RF magnetic field is applied along the width of the ferromagnetic layers.

To study the effect of the air gap thickness on power loss due to eddy currents, a simple structure of two conductive ferromagnetic layers separated by an air gap (Fig. A1) is used. By Faraday's Law, the RF magnetic field applied to the sample will generate an electromotive force, which will cause eddy currents to circulate in the ferromagnetic layers. When the gap between the ferromagnetic layers is small, the two layers become strongly capacitively coupled and displacement currents travel between them, leading to an increase in power dissipation as new eddy current paths are created [23]–[26]. If the air gap is too thin, the eddy current suppression effects of the laminations becomes practically nonexistent.

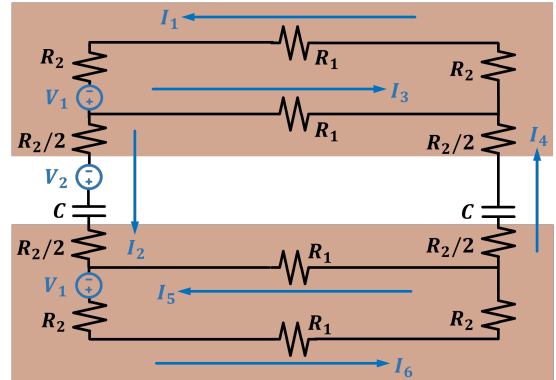

Fig. A2. Equivalent circuit model used to intuitively study eddy currents generated in the structure depicted in Fig. A1.

To analyze the structure in Fig. A1, the approach of [27] is used and the structure is modelled by an equivalent circuit (Fig. A2). Here, the resistive elements $R_1$ and $R_2$ represents the power dissipation by eddy currents traveling along the length and thickness, respectively. The capacitive elements are used to account for the coupling between the layers, and the voltage sources account for the electromotive force generated by the RF magnetic field through Faraday's Law. The values for the lumped elements and voltage sources used in the circuit model are shown in Table 1. Additional interlayer coupling effects (such as dipolar magnetic fields between the layers), dynamic magnetization effects due to ferromagnetic resonance, or incorporating a large number of layers would be difficult to capture using this approach and is a task better suited for the ADI-FDTD model discussed in the main body of this work.

| Variable | Formula | Description |
|---|---|---|
| $\delta$ | $\sqrt{2/(\sigma\mu_r\mu_0\omega)}$ | Skin depth in ferromagnetic layer |
| $d_1$ | $\delta \cdot \{1 - \exp[-t_m/(2\delta)]\}$ | Effective depth of eddy currents traveling along the length of the structure |
| $d_2$ | $\delta \cdot \{1 - \exp[-l/(2\delta)]\}$ | Effective depth of eddy currents traveling along the thickness of the structure |
| $C$ | $\epsilon_0 lw/(2t_2)$ | Capacitance coupling the eddy currents between the two ferromagnetic layers |
| $R_1$ | $l/(\sigma w d_1)$ | Approximate resistance seen by eddy currents traveling along the length of the structure |
| $R_2$ | $t_1/(2\sigma w d_2)$ | Approximate resistance seen by eddy currents traveling along the thickness of the structure |
| $Z_c$ | $1/(j\omega \cdot C)$ | Impedance due to capacitive coupling of eddy currents |
| $V_1$ | $\omega \cdot \mu_r\mu_0 \cdot l \cdot (t_1/2) \cdot H_{RF}$ | Electromotive force acting on eddy current loops contained solely in the ferromagnetic layer |
| $V_2$ | $V_1 + (\omega \cdot \mu_0 \cdot l \cdot t_2 \cdot H_{RF})$ | Electromotive force acting on eddy current loops crossing between the adjacent two layers |

Table 1: Formulas for variables used to calculate equivalent circuit parameters. Note that in the formulas for $R_1$ and $R_2$, crowding of the current due to the skin effect is considered through the parameters $d_1$ and $d_2$ [27].

### B. Determining Unknown Currents

There are six unknown currents in the circuit shown in Fig. A2 that must be solved for to calculate the total power dissipated by the eddy currents. Using Kirchhoff's voltage and current laws, six equations can be derived to solve for the six unknowns.

$$V_1 - 2I_1 R_2 - (I_1 + I_3)R_1 = 0, \tag{A1a}$$

$$V_2 - (I_2 + I_4) \cdot (R_2 + Z_c) + (I_3 + I_5)R_1 = 0, \tag{A1b}$$

$$V_1 - 2I_6 R_2 - (I_6 + I_5)R_1 = 0, \tag{A1c}$$

$$I_1 - I_3 = I_2 , \tag{A1d}$$

$$I_2 + I_5 = I_6 , \tag{A1e}$$

$$I_4 + I_3 = I_1 , \tag{A1f}$$

By adding Equations (A1d) and (A1f), we can solve for $I_4$.

$$(I_1 - I_3) + (I_4 + I_3) = I_2 + I_1$$

$$I_4 = I_2 \tag{A2a}$$

Using Equation (A2a), Equations (A1e) and (A1f) can be subtracted from each other to get the following relation.

$$(I_2 + I_5) - (I_2 + I_3) = I_6 - I_1$$

$$I_5 - I_3 = I_6 - I_1 \tag{A2b}$$

Rewriting Equations (A1a) and (A1c), solutions for $I_3$ and $I_5$ can be found.

$$I_3 = \frac{V_1}{R_1} - \left(1 + 2\frac{R_2}{R_1}\right)I_1 \tag{A2c}$$

$$I_5 = \frac{V_1}{R_1} - \left(1 + 2\frac{R_2}{R_1}\right)I_6 \tag{A2d}$$

Combining Equations (A2b), (A2c), and (A2d), $I_1$ and $I_6$ can be solved for.

$$-\left(1 + 2\frac{R_2}{R_1}\right)I_6 + \left(1 + 2\frac{R_2}{R_1}\right)I_1 = I_6 - I_1$$

$$\left(1 + 2\frac{R_2}{R_1}\right)(I_6 - I_1) = I_6 - I_1 \tag{A2e}$$

The only way Equation (A2e) can be true for any value of $\frac{R_2}{R_1}$ is if $I_6 - I_1$ is zero. Combining this with Equation (A2b), gives the following two equations for $I_5$ and $I_6$.

$$I_5 = I_3 \tag{A2f}$$

$$I_6 = I_1 \tag{A2g}$$

With Equations (A2a), (A2f), and (A2g) in mind, the currents in Fig. A2 can be relabeled to reduce the number of unknowns down to three.

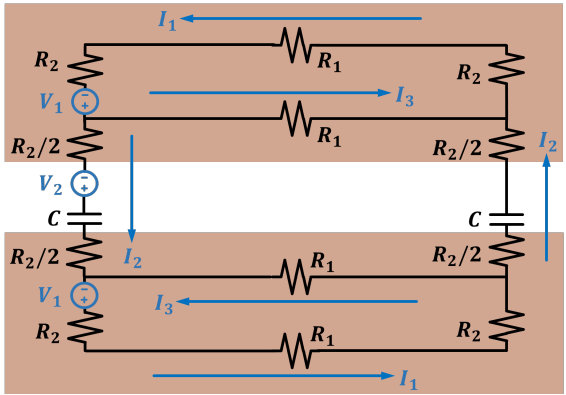

Fig. A3. Equivalent circuit model shown in Fig. A2, but with Equations (A2a), (A2f), and (A2g) applied to reduce the number of unknowns.

Rewriting Equations (A1a) and (A1b), gives the following two equations for the remaining three currents.

$$\frac{V_1}{R_1} - \left(1 + 2\frac{R_2}{R_1}\right)I_1 - I_3 = 0 \tag{A3a}$$

$$\frac{V_2}{2R_1} - \left(\frac{R_2}{R_1} + \frac{Z_c}{R_1}\right)I_2 + I_3 = 0 \tag{A3b}$$

Combining Equation (A1d), (A3a), and (A3b), $I_3$ can be eliminated and two equations for $I_1$ and $I_2$ can be found.

$$\frac{V_1}{R_1} - \left(1 + 2\frac{R_2}{R_1}\right)I_1 - I_1 + I_2 = 0$$

$$\frac{V_1}{R_1} - 2\left(1 + \frac{R_2}{R_1}\right)I_1 + I_2 = 0 \tag{A3c}$$

$$\frac{V_2}{2R_1} - \left(\frac{R_2}{R_1} + \frac{Z_c}{R_1}\right)I_2 + I_1 - I_2 = 0$$

$$\frac{V_2}{2R_1} - \left(1 + \frac{R_2}{R_1} + \frac{Z_c}{R_1}\right)I_2 + I_1 = 0 \tag{A3d}$$

Using Equations (A1d), (A3c), and (A3d), $I_1$, $I_2$, and $I_3$ can readily be solved.

$$D = 2\left(1 + \frac{R_2}{R_1}\right)\left(1 + \frac{R_2}{R_1} + \frac{Z_c}{R_1}\right) - 1 \tag{A4a}$$

$$I_1 = \frac{\frac{V_1}{R_1}\left(1 + \frac{R_2}{R_1} + \frac{Z_c}{R_1}\right) + \frac{V_2}{2R_1}}{D} \tag{A4b}$$

$$I_2 = \frac{\frac{V_1}{R_1} + \left(1 + \frac{R_2}{R_1}\right)\frac{V_2}{R_1}}{D} \tag{A4c}$$

$$I_3 = \frac{\frac{V_1}{R_1}\left(\frac{R_2}{R_1} + \frac{Z_c}{R_1}\right) - \frac{V_2}{R_1}\left(\frac{1}{2} + \frac{R_2}{R_1}\right)}{D} \tag{A4d}$$

## C. Perfect Insulation Limit

Assuming $R_1 \gg R_2$, in the limit of a perfectly insulating air gap ($Z_c \to \infty$), these equations reduce down to the following equations.

$$I_1 = \frac{V_1}{2R_1} \tag{A5a}$$

$$I_2 = 0 \tag{A5b}$$

$$I_3 = \frac{V_1}{2R_1} \tag{A5c}$$

In this limit, there is no coupling between the layers and the entirety of the eddy currents are contained within the individual ferromagnetic layers (Fig. A4a). Power dissipation in this scenario will be the minimum possible for the structure shown in Fig. A1.

## D. No Air Gap Limit

Again assuming $R_1 \gg R_2$, in the limit where the air gap goes to zero ($t_{air} \to 0$, $V_2 \to V_1$, $Z_c \to 0$), Equations (A4b), (A4c), and (A4d) reduce down to the following equations.

$$I_1 = \frac{3}{2}\frac{V_1}{R_1} \tag{A6a}$$

$$I_2 = 2\frac{V_1}{R_1} \tag{A6b}$$

$$I_3 = -\frac{V_1}{2R_1} \tag{A6c}$$

In this limit, $I_3$ now reverses direction as additional eddy currents are allowed to flow throughout the structure since the two ferromagnetic layers are now shorted together (Fig. A4(b)). Power dissipation in this scenario is now the maximum possible for the structure shown in Fig. A1.

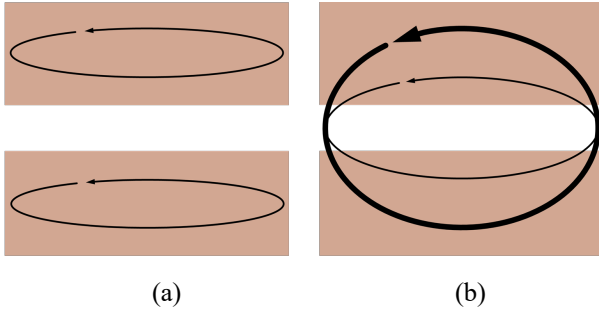

|                    (a)                    |                    (b)                    |

Fig. A4. (a) Current flow in the perfect insulation limit. (b) Current flow in the no air gap limit.

### E. Power Dissipation vs Air Gap Thickness

The total power dissipated due to eddy currents is simply the sum of the power dissipated in each of the resistors in the equivalent circuit. Note that because of the reactive contribution from capacitive coupling between the layers, the currents found through Equations (A4b), (A4c), and (A4d) are generally complex.

$$P_{dissipated} = 2 \times \left(\frac{1}{2}I_1 I_1^* \cdot (R_1 + 2R_2) + \frac{1}{2}I_2 I_2^* \right.$$
$$\left. \cdot R_2 + \frac{1}{2}I_3 I_3^* \cdot R_1\right) \tag{A7}$$

Equation (A7) is a strong function of $t_{air}$, as it not only tunes the capacitive coupling between the layers ($Z_c$), but also the electromotive force felt by the central current loop ($V_2$). For small thickness, the power dissipation will increase as it approaches the limit of completely shorted ferromagnetic layers. As the thickness increases, the impedance of the air gap also increases and eventually approaches the limit of the perfectly insulating air gap.

Shown below is Equation (A7) plotted as a function of air gap thickness. Here, the structure has in-plane dimensions of $300\ \mu\text{m} \times 300\ \mu\text{m}$ and the two ferromagnetic layers are $0.24\ \mu\text{m}$ thick. The ferromagnetic material is FeGaB at ferromagnetic resonance (2.4 GHz for a DC bias of 60 Oe). The plot is normalized to the power dissipated in the no air gap limit.

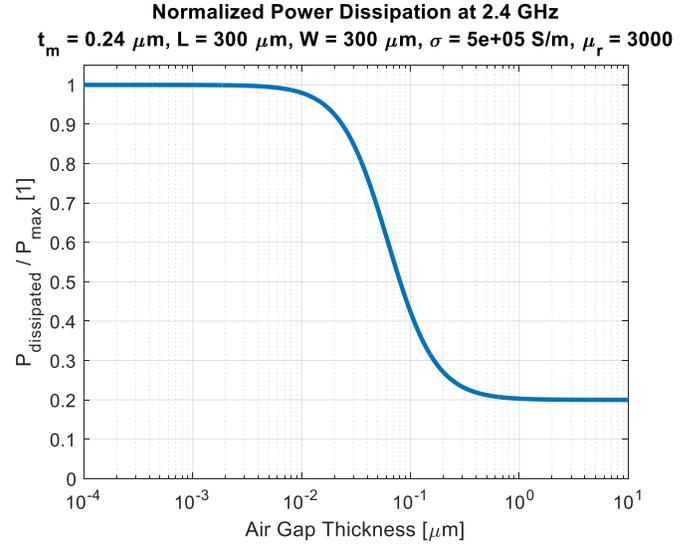

**Normalized Power Dissipation at 2.4 GHz**
$t_m$ = 0.24 $\mu$m, L = 300 $\mu$m, W = 300 $\mu$m, $\sigma$ = 5e+05 S/m, $\mu_r$ = 3000

Fig. A5. Power dissipation for the structure in Fig. A1 as a function of air gap thickness, where the ferromagnetic material is FeGaB at ferromagnetic resonance.